\documentclass[nofootinbib,twocolumn,aps,letterpaper,superscriptaddress,showpacs]{revtex4}
\usepackage{amsmath}
\usepackage{amsfonts}
\usepackage[dvips]{graphicx}
\begin{document}

\newcommand\be{\begin{equation}}
\newcommand\ee{\end{equation}}
\newcommand\bea{\begin{eqnarray}}
\newcommand\eea{\end{eqnarray}}
\newcommand\bseq{\begin{subequations}} %solo con amsmath
\newcommand\eseq{\end{subequations}}
\newcommand\bcas{\begin{cases}}
\newcommand\ecas{\end{cases}}
\newcommand{\p}{\partial}
\newcommand{\f}{\frac}

\title{Semiclassical suppression of weak anisotropies of a generic Universe}

\author{Marco Valerio Battisti}
\email{battisti@icra.it}
\affiliation{ICRA - International Center for Relativistic Astrophysics}
\affiliation{Centre de Physique Th\'eorique de Luminy, Universit\'e de la M\'editerran\'ee F-13288 Marseille, FR}
\affiliation{Dipartimento di Fisica (G9), Universit\`a di Roma ``Sapienza'' P.le A. Moro 5 00185 Rome, IT}
\author{Riccardo Belvedere}
\email{riccardo.belvedere@icra.it}
\affiliation{ICRA - International Center for Relativistic Astrophysics}
\affiliation{Dipartimento di Fisica (G9), Universit\`a di Roma ``Sapienza'' P.le A. Moro 5 00185 Rome, IT}
\author{Giovanni Montani}
\email{montani@icra.it} 
\affiliation{ICRA - International Center for Relativistic Astrophysics}
\affiliation{Dipartimento di Fisica (G9), Universit\`a di Roma ``Sapienza'' P.le A. Moro 5 00185 Rome, IT}
\affiliation{ENEA C.R. Frascati (Dipartimento FPN), Via E. Fermi 45 00044 Frascati Rome, IT}
\affiliation{ICRANET C.C. Pescara, P.le della Repubblica 10 65100 Pescara, IT}

%\today

\begin{abstract}
A semiclassical mechanism which suppresses the weak anisotropies of an inhomogeneous cosmological model is developed. In particular, a wave function of this Universe having a meaningful probabilistic interpretation is obtained that is in agreement with the Copenhagen School. It describes the evolution of the anisotropies with respect to the isotropic scale factor which is regarded as a semiclassical variable playing an observer-like role. Near the cosmological singularity the solution spreads over all values of the anisotropies while, when the Universe expands sufficiently, the closed Friedmann-Robertson-Walker model appears to be the favorite state.
\end{abstract}

\pacs{98.80.Qc; 04.60.Ds; 04.60.Bc}

\maketitle 

\section{Introduction}

Quantum cosmology denotes the application of the quantum theory to the entire Universe as described by cosmological models (for reviews see \cite{QC}). All but a finite number of degrees of freedom are frozen out by imposing symmetries (homogeneity or isotropy) and the resulting finite dimensional configuration space of the theory is known as minisuperspace. The resulting framework is thus a natural arena to test ideas and constructions introduced in the (not yet found) quantum theory of gravity. 

In quantum cosmology the Universe is described by a single wave function $\Psi$ providing puzzling interpretations as soon as the differences with respect to ordinary quantum mechanics are addressed \cite{Vil,pqc} (see also \cite{Pin}). Quantum cosmology is defined up to the following two assumptions. (i) The analyzed model is the Universe as a whole and thus there is no longer an a priori splitting between classical and quantum worlds. No external measurement crutch is available and an internal one can not play the observer-like role because of the extreme conditions a primordial Universe is subjected to. (ii) In general relativity time is an arbitrary label and clocks, being parts of the Universe, are also described by the wave function $\Psi$. Time is thus included in the configuration space and the integral of $|\Psi|^2$ over the whole minisuperspace diverges as in quantum mechanics when the time coordinate is included in the configuration-space element. As a result the standard interpretation of quantum mechanics (the Copenhagen interpretation) does not work in quantum cosmology. On a given (space-time) background structure only, observations can take place in the sense of ordinary quantum theory. 

In this work a wave function of a generic inhomogeneous Universe, which has a clear probabilistic interpretation, is obtained. It can be meaningfully interpreted because of a separation between semiclassical degrees of freedom, in the Wentzel-Kramers-Brillouin (WKB) sense, and quantum ones. In particular, the quantum dynamics of weak anisotropies (the physical degrees of freedom of the Universe) is traced with respect to the isotropic scale factor which plays an observer-like role as soon as the Universe expands sufficiently. 

A generic inhomogeneous cosmological model, describing a Universe in which any specific symmetry has been removed, represents a generic cosmological solution of the Einstein field equations \cite{BKL}. Belinski-Khalatnikov-Lifshitz (BKL) showed that such a geometry evolves asymptotically to the singularity as an ensemble, one for each causal horizon, of independent Bianchi IX homogeneous Universes \cite{Mis}. This model represents the best description we have of the (classical) physics near a space-like cosmological singularity. 

Our main result is that the wave function of the Universe is spread over all values of anisotropies near the cosmological singularity, but it is asymptotically peaked around the isotropic configuration. The closed Friedmann-Robertson-Walker (FRW) cosmological model is then the naturally privileged state as soon as a sufficiently large volume of the Universe is taken into account. A semiclassical isotropization mechanism for the Universe is thus predicted.

This model can be regarded as a concrete implementation, to a physically interesting cosmological problem, of the semiclassical approach to quantum cosmology \cite{Vil}. An isotropization mechanism is in fact necessary to explain the transition between a very early Universe and the observed one. The isotropic FRW model can accurately describes the evolution of the Universe until decoupling time, i.e. until $10^{-3}-10^{-2} s$ after the big-bang \cite{Kolb}. On the other hand, the description of its primordial stages requires more general models. It is thus fundamental to recover a mechanism which can match these two cosmological epochs. Although many efforts have been made inside classical theory \cite{KM,iso} (especially by the use of the inflation field), no quasiclassical (or purely quantum) isotropization mechanism is yet developed in detail (for different attempts see \cite{qiso}.) In particular this work improves \cite{MR} in which no clear (unique) probabilistic interpretation of the wave function can be formulated at all.  

The paper is organized as follows. In Section II the wave function of a generic Universe and its interpretation are analyzed. Section III is devoted to the study of an isotropization mechanism. Finally, in Section IV the validity of the model is discussed. Concluding remarks follows.

We adopt natural units $\hbar=c=1$ apart from where the classical limits are discussed. 

\section{Wave function in quantum cosmology}

In this Section a wave function of a generic inhomogeneous Universe, which has a meaningful probabilistic interpretation, is described. There are some reliable indications that the early stages of the Universe evolution are characterized by such a degree of generality \cite{KM,ben}. In the quantum regime, dealing with the absence of global symmetry, it is however required by indeterminism. On different causal regions the geometry has to fluctuate independently so preventing global isometries. 

The dynamics of a generic cosmological model is summarized, asymptotically to the cosmological singularity, in the action $I=\int_{\mathcal M} dtd^3x\left(p_i\p_tq^i-N\mathcal H\right)$ \cite{ben} (for a review see \cite{rev}). Here $q^i$ are the three scale factors, $p_i$ the three conjugate momenta, $N$ the lapse function and $\mathcal H=0$ is the scalar constraint which, in the Misner scheme, reads
\be\label{scacon} 
\mathcal H(x^i)=\kappa\left[-\f{p_a^2}a+\f1{a^3}\left(p_+^2+p_-^2\right)\right]+\f{a}{4\kappa}V(\beta_\pm)+U(a)=0.
\ee  
The potential term $V(\beta_\pm)$ accounts for the spatial curvature of the model and is given by
\begin{multline}\label{pmix}
V=\lambda_1^2\left(e^{-8\beta_+}-2e^{4\beta_+}\right)+\lambda_2^2\left(e^{4(\beta_++\sqrt3 \beta_-)}-2e^{-2(\beta_++}\right.\\\left.^{-\sqrt3\beta_-)}\right)+\lambda_3^2\left(e^{4(\beta_+-\sqrt3\beta_-)}-2e^{-2(\beta_++\sqrt3 \beta_-)}\right)+\lambda_{in}^2.
\end{multline}
In this expression $l_{in}^i=a/\lambda_i$ ($\lambda_i=\lambda_i(x^i)$) denotes the co-moving physical scale of inhomogeneities \cite{BM06}, $\lambda^2=\lambda_1^2+\lambda_2^2+\lambda_3^2$ and $\kappa=8\pi G$ is the Einstein constant. The function $a=a(t,x^i)$ describes the isotropic expansion of the Universe while its shape changes (the anisotropies) are associated to $\beta_\pm=\beta_\pm(t,x^i)$. It is relevant to remark an important feature of such a model. Via the BKL scenario \cite{BKL} the dynamics of a generic inhomogeneous Universe reduces, point by point, to the one of a Bianchi IX model. More precisely the spatial points dynamically decouple toward the singularity and a generic Universe is thus described by a collection of causal regions each of which evolves independently as a homogeneous model, in general as Bianchi IX. This picture holds as far as the inhomogeneities are stepped out of the cosmological horizon $l_h\sim t$, i.e. the inequality $l_{in}\gg l_h$ holds \cite{BKL}. In each space point the phase space is then six dimensional with coordinates $(a,p_a,\beta_\pm,p_\pm)$ and the cosmological singularity appears as $a\rightarrow0$. In this system the matter terms are regarded negligible with respect to the cosmological constant, i.e. the isotropic potential $U$ in (\ref{scacon}) reads 
\be
U(a)=-\f{a\lambda^2}{4\kappa}+\f{\Lambda a^3}\kappa
\ee
$\Lambda$ being the cosmological constant. Far enough from the singularity the cosmological constant term dominates on the ordinary matter fields. Such a contribution is necessary in order for the inflationary scenario to take place \cite{KM,iso,Kolb}. 

As we said, a correct definition of probability (positive semidefinite) in quantum cosmology can be formulated by distinguishing between semiclassical and quantum variables \cite{Vil}. Variables satisfying the Hamilton-Jacobi equation are regarded as semiclassical and it is assumed that the semiclassical dynamics is not affected by quantum dynamics. In this respect the quantum variables describe a small subsystem of the Universe. It is thus natural to regard the isotropic scale factor $a$ as semiclassical and to consider the anisotropies $\beta_\pm$ (the two physical degrees of freedom of the Universe) as quantum variables. In other words, we assume ab initio that the radius of the Universe is of different nature with respect to its shape changes. As we will see the isotropic share of the scalar constraint becomes semiclassical before the anisotropic one. In agreement with such a reasoning the wave functional of the Universe $\Psi=\Psi(a,\beta_\pm)$ reads \cite{Vil} 
\be\label{wvil}
\Psi\stackrel{a\rightarrow0}\longrightarrow\prod_i\Psi_i(x^i), \quad \Psi_i=\psi_0\chi=A(a)e^{iS(a)}\chi(a,\beta_\pm)
\ee 
where the factorization is due to decoupling of the spatial point. This wave function is WKB-like in $a$ (amplitude and phase depend only on the semiclassical variable). The additional function $\chi$ depends on the quantum variables $\beta_\pm$ and parametrically only, in the sense of the Born-Oppenheimer approximation, on the scale factor. The effects of the anisotropies on the Universe expansion, as well as the effects of the electrons on the dynamics of nuclei, are thus regarded as negligible.

The canonical quantization of this model is achieved by the use of the Dirac prescription for quantizing constrained systems \cite{HT}, i.e. imposing that the physical states are those annihilated by the self-adjoint operator $\hat{\mathcal H}$ corresponding to the classical counterpart (\ref{scacon}). We represent the minisuperspace canonical commutation relations $[\hat q_i,\hat p_j]=i\delta_{ij}$ in the coordinate space, i.e. $\hat q_i$ and $\hat p_i$ act as multiplicative and derivative operators respectively. The Wheeler-DeWitt (WDW) equation for this model leads, considering (\ref{wvil}), to three different equations. We obtain the Hamilton-Jacobi equation for $S$ and the equation of motion for $A$, which respectively read
\be\label{hj}
-\kappa A\left(S'\right)^2+aUA+\mathcal V_q=0, \qquad \f1A\left(A^2S'\right)'=0.
\ee
Here $(\cdot)'=\p_a$ and $\mathcal V_q=\kappa A''$ is the so-called quantum potential which is negligible far from the singularity even if the $\hbar\rightarrow0$ limit is not taken into account (see below). The action $S(a)$ defines a congruence of classical trajectories, while the second equation in (\ref{hj}) is the continuity equation for the amplitude $A(a)$.

The third equation we achieve describes the evolution of the quantum subsystem and is given by
\be\label{weq}
a^2\left(2A'\p_a\chi+A\p_a^2\chi+2iAS'\p_a\chi\right)+A\hat H_q\chi=0,
\ee   
where 
\be
H_q=p_+^2+p_-^2+\f{a^4}{4\kappa^2}V(\beta_\pm)
\ee
represents the quantum Hamiltonian. The first two terms in (\ref{weq}) are of higher order in $\hbar$ than the third and can be neglected. We then deal with a Schr\"odinger-like equation for the quantum wave function $\chi$ 
\be\label{eveq}
-2ia^2S'\p_a\chi=\hat H_q\chi.
\ee
Such an equation is in agreement with the assumption that the anisotropies describe a quantum subsystem of the whole Universe, i.e. that the wave function $\chi$ depends on $\beta_\pm$ only (in the Born-Oppenheimer sense). The smallness of such a quantum subsystem can be formulated requiring that its Hamiltonian $H_q$ is of order $\mathcal O(\epsilon^{-1})$, where $\epsilon$ is a small parameter proportional to $\hbar$. Since the action of the semiclassical Hamiltonian operator $\hat H_0=a^2\p_a^2+a^3U/\kappa$ on the wave function $\Psi_i$ is of order $\mathcal O(\epsilon^{-2})$, the idea that the anisotropies do not influence the isotropic expansion of the Universe can be formulated as $\hat H_q\Psi_i/\hat H_0\Psi_i=\mathcal O(\epsilon)$. Such a requirement is physically reasonable since the semiclassical properties of the Universe, as well as the smallness of the quantum subsystem, are both related to the fact that the Universe is sufficiently large \cite{Vil}.  

A pure Schr\"odinger equation for $\chi$ is obtained taking into account the tangent vector to the classical path. From the first of equations (\ref{hj}) we find
\be\label{pa}
p_a=S'=-\f a\kappa\sqrt{\Lambda a^2-\f{\lambda^2}4},
\ee
and the minus sign is chosen to have compatibility between the time gauge $da/dt=1$ and a positive lapse function $2N=(\Lambda a^2-\lambda^2/4)^{-1/2}$. It is then possible to define a new time variable $\tau$ such that $d\tau=(N\kappa/a^3)da$ and, considering the lapse function $N$, it reads
\be\label{tim}
\tau=\f\kappa{a^2}\left[\f{\sqrt{4\Lambda a^2-\lambda^2}}2-2\Lambda a^2\tan^{-1}\left(\f1{\sqrt{4\Lambda a^2-\lambda^2}}\right)\right].
\ee 
This equation can be simplified in the asymptotic region $a\gg\lambda/\sqrt\Lambda$ where $\tau$ behaves as $\tau=(\kappa/12\sqrt\Lambda)a^{-3}+\mathcal O(a^{-5})$. Such a region deserves interest since the variable $a$ can be considered as semiclassical and  an isotropization mechanics for the Universe takes place (see below). Choosing $\tau$ as time coordinate, equation (\ref{eveq}) rewrites as
\be\label{scheq}
i\p_\tau\chi=\hat H_q\chi=\left(-\Delta_\beta+\f{a^4}{4\kappa^2}V(\beta_\pm)\right)\chi,
\ee
which has the desired form. 

Let us now discuss the implications of this approach for the definition of the probability distribution. The wave function (\ref{wvil}) defines a probability distribution $\rho(a,\beta_\pm)=\rho_0(a)\rho_\chi(a,\beta_\pm)$, where $\rho_0(a)$ is the classical probability distribution for the semiclassical variable $a$. On the other hand, $\rho_\chi=|\chi|^2$ denotes the probability distribution for the quantum variables $\beta_\pm$ on the classical trajectories (\ref{hj}) where the wave function $\chi$ can be normalized. An ordinary interpretation (in the Copenhagen sense) of a wave function tracing a subsystem of the Universe is therefore recovered.

\section{The isotropization mechanism}

A wave function which naturally leads to an isotropic configuration of the Universe is here obtained. As we said the generic inhomogeneous cosmological model is described, toward the singularity, by a collection of $\infty^3$ independent Bianchi IX models each of which referred to a different spatial point \cite{BKL}. Bianchi IX (the Mixmaster Universe \cite{Mis}) is the most general, together with Bianchi VIII, homogeneous cosmological model and its spatial geometry is invariant under the $SO(3)$ group \cite{rev,RS,chaos}. This system generalizes the closed FRW cosmological dynamics if the isotropy hypothesis is relaxed.  

In order to enforce the idea that the anisotropies can be considered as the only quantum degrees of freedom of the Universe we address the quasi-isotropic regime, i.e. $|\beta_\pm|\ll1$. Moreover, since we are interested in the link existing between the isotropic and anisotropic dynamics, the Universe has to be get through to such a quasi-isotropic era. In this regime the potential term reads $V(\beta_\pm)=8\lambda^2(\beta_+^2+\beta_-^2)+\mathcal O(\beta^3)$ while for the equation (\ref{scheq}) we get 
\be\label{osceq}
i\p_\tau\chi=\f12\left(-\Delta_\beta+\omega^2(\tau)(\beta_+^2+\beta_-^2)\right)\chi,
\ee
where $\tau$ has been rescaled by a factor $2$ and $\omega^2(\tau)=C/\tau^{4/3}$ ($C$ being a constant in each space point given by $2C=\lambda^2((6)^{4/3}(\kappa\Lambda)^{2/3})^{-1}$). The dynamics of the Universe anisotropies subsystem can then be regarded as a time-dependent bi-dimensional harmonic oscillator with frequency $\omega(\tau)$. 

The construction of the quantum theory for a time-dependent, linear, dynamical system has remarkable differences with respect to the time-independent one \cite{Wald}. If the Hamiltonian fails to be time-independent, solutions which oscillate with purely positive frequency do not exist at all (the dynamics is not carried out by an unitary time operator). In particular, in the absence of a time translation symmetry, no natural preferred choice of the Hilbert space is available. In the finite-dimensional case (where the Stone-Von Neumann theorem holds) no real mistake arises since for any choice of the Hilbert space the theory is unitarily equivalent to the standard one. On the other hand, as soon as a field theory is taken into account serious difficulties appear and an algebraic approach is required \cite{Wald}. 
  
The quantum theory of a harmonic oscillator with time-dependent frequency is known and the solution of the Schr\"odinger equation can be obtained analytically \cite{harosc}. The analysis is mainly based on the use of the ``exact invariants method'' and on some time-dependent transformations. An exact invariant $J(\tau)$ is a constant of motion (namely $\dot J\equiv dJ/d\tau=\p_\tau J-i[J,\hat H_q]=0$), is hermitian ($J^\dag=J$) and for the Hamiltonian $H_q$ as in (\ref{osceq}) it explicitly reads
\be\label{inv} 
J_\pm=\f12\left(\rho^{-2}\beta_\pm^2+(\rho p_\pm-\dot\rho\beta_\pm)^2\right).
\ee
Here $\rho=\rho(\tau)$ is any function satisfying the auxiliary non-linear differential equation $\ddot\rho+\omega^2\rho=\rho^{-3}$. The goal for the use of such invariants (\ref{inv}) relies on the fact that they match the wave function of a time-independent harmonic oscillator with the time-dependent one. Let $\phi_n(\beta,\tau)$ the eigenfunctions of $J$ forming a complete orthonormal set corresponding to the time-independent eigenvalues $k_n=n+1/2$. These states turns out to be related to the eigenfunctions $\tilde\phi_n(\xi)$ ($\xi=\beta/\rho$) of a time-independent harmonic oscillator via the unitary transformation $U=\exp(-i\dot\rho\beta^2/2\rho)$ as $\tilde\phi_n=\rho^{1/2} U\phi_n$. The non-trivial (and in general non-available) step in this construction is to obtain an exact solution of the auxiliary equation for $\rho$. However in our case it can be constructed and explicitly reads
\be\label{rho}
\rho=\sqrt{\f\tau{\sqrt C}\left(1+\f{\tau^{-2/3}}{9C}\right)}.
\ee
Finally, the solution of the Schr\"odinger equation (\ref{osceq}) is connected to the $J$-eigenfunctions $\phi_n$ by the relation $\chi_n(\beta,\tau)=e^{i\alpha_n(\tau)}\phi_n(\beta,\tau)$. (The general solution of (\ref{osceq}) can thus be written as the linear combination $\chi(\beta,\tau)=\sum_nc_n\chi_n(\beta,\tau)$, $c_n$ being constants.) Here the time-dependent phase $\alpha_n(\tau)$ is given by
\begin{multline}\label{alp}
\alpha_n=-\left(n+\f12\right)\int\f{d\tau}{\rho^2(\tau)}=\\=\f{3\sqrt C}2\left(n+\f12\right)\left[\ln(9C^{3/2})+\ln\left(\f1{9C^{3/2}}+\f{\tau^{2/3}}{\sqrt C}\right)\right].
\end{multline}     
Collecting these results the wave function $\chi_n(\beta_\pm,\tau)$, which describes the evolution of the anisotropies of the Universe with respect to the scale factor, reads $\chi_n=\chi_{n_+}(\beta_+,\tau)\chi_{n_-}(\beta_-,\tau)$. Here $n=n_++n_-$ and
\be\label{chi}
\chi_{n_\pm}(\beta_\pm,\tau)=A\f{e^{i\alpha_n(\tau)}}{\sqrt\rho}h_n(\xi_\pm)\exp\left[\f i2\left(\dot\rho\rho^{-1}+i\rho^{-2}\right)\beta_\pm^2\right],
\ee
where $A$ is the normalization constant, $h_n$ are the usual Hermite polynomial of order $n$ and $\rho(\tau)$ and the phase $\alpha(\tau)$ are given by (\ref{rho}) and (\ref{alp}), respectively. It is immediate to verify that, as $\omega(\tau)\rightarrow\omega_0$ and $\rho(\tau)\rightarrow\rho_0=1/\sqrt{\omega_0}$ (namely $\alpha(\tau)\rightarrow-\omega_0k_n\tau$), the solution of the time-independent harmonic oscillator is recovered.

Let us investigate the probability density to find the quantum subsystem of the Universe at a given state. As a result the anisotropies appear to be probabilistically suppressed as soon as the Universe expands sufficiently far from the cosmological singularity (it appears for $a\rightarrow0$ or $\tau\rightarrow\infty$). Such a feature can be realized from the behavior of the squared modulus of the wave function (\ref{chi}) which is given by
\be\label{prob}
|\chi_n|^2\propto\f1{\rho^2}|h_{n_+}(\xi_+)|^2|h_{n_-}(\xi_-)|^2 e^{-\beta^2/\rho^2},
\ee  
where $\beta^2=\beta^2_++\beta^2_-$. This probability density is still time-dependent through $\rho=\rho(\tau)$ and $\xi_\pm=\beta_\pm/\rho$ since the evolution of the wave function $\chi$ is not traced by an unitary time operator. As we can see from (\ref{prob}) when a large enough isotropic cosmological region is considered (namely when the limit $a\rightarrow\infty$ or $\tau\rightarrow0$ is taken into account) the probability density to find the Universe is sharply peaked around the isotropic configuration $\beta_\pm=0$. In this limit (which corresponds to $\rho\rightarrow0$) the probability density $|\chi_{n=0}|^2$ of the ground state $n=0$ is given by  $|\chi_{n=0}|^2\stackrel{\tau\rightarrow0}\longrightarrow\delta(\beta,0)$. It is thus proportional to the Dirac $\delta$-distribution centered on $\beta=0$ (see Fig. 1). 
\begin{figure}
\begin{center}
\includegraphics[height=2.3in]{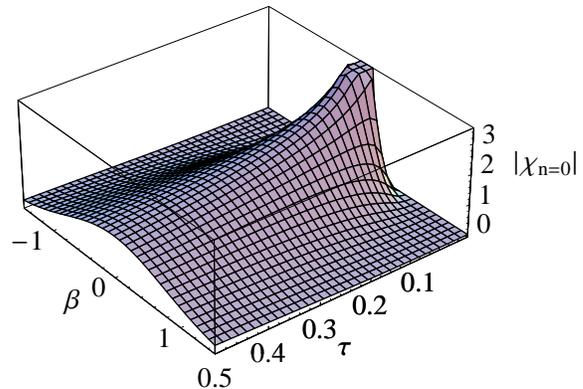}
\caption{The absolute value of the ground state of the wave function $\chi(\beta_\pm,\tau)$ far from the cosmological singularity. In the plot we take $C=1$.} 
\end{center}
\end{figure} 

Summarizing, when the Universe moves away from the cosmological singularity, the probability density to find it is asymptotically peaked (as a Dirac $\delta$-distribution) around the closed FRW configuration. Near the singularity all values of the anisotropies $\beta_\pm$ are almost equally favored from a probabilistic point of view. On the other hand, as the volume of the Universe grows, the isotropic state becomes the most probable state of the Universe.

The key feature of such a result relies on the fact that the isotropic  scalar factor has been considered as an intrinsic variable with respect to the anisotropies. It has been treated semiclassically (WKB) while the two physical degrees of freedom of the Universe ($\beta_\pm$) have been described as real quantum coordinates (the validity of this assumption is discussed in what follows). In this way a positive semidefinite probability density can be constructed for the wave function of the quantum subsystem of the Universe.

\section{Physical considerations on the model}

To complete our analysis we investigate the range of validity of the model. In particular, (i) we want to analyze in which sense it is correct to regard the scale factor $a$ as a WKB variable and (ii) up to which regime the hypothesis of a quasi-isotropic potential is reasonable. 

A variable can be considered semiclassical in the WKB sense if its dynamics is completely described by the zero-order Hamilton-Jacobi equation \cite{Sak}. If the dynamics of a system no longer evolves according to the Schr\"odinger equation but all the informations are summarized in the Hamilton-Jacobi equation, then such a system can be regarded as classical. In order to understand in which regime the Hamiltonian $H_0=-a^2p_a^2+a^3U/\kappa$ is semiclassical (or equivalently when $a$ is a WKB variable) the roles of the quantum potential $\mathcal V_q$ and the WKB wave function $\psi_0$ have to be investigated. The term $\mathcal V_q$ in the Hamilton-Jacobi equation (\ref{hj}) can be easily computed. From the continuity equation (the second of (\ref{hj})) and (\ref{pa}), the amplitude $A$ turns out to be 
\be\label{A}
A=\sqrt{\f{\sqrt\kappa}a}\left(\Lambda a^2-\f{\lambda^2}4\right)^{-1/4}.
\ee
In the $a\gg\lambda/\sqrt\Lambda$ limit ($\tau\rightarrow0$) the quantum potential behaves like $\mathcal V_q\sim\mathcal O(1/a^3)$. In other words $\mathcal V_q$ can be neglected as soon as the Universe sufficiently expands, even if the limit $\hbar\rightarrow0$ is not taken into account. It is also easy to verify that the function $\psi_0=\exp(iS+\ln A)$ approaches the quasi-classical limit $e^{iS}$ for sufficiently large values of $a$. In fact, considering (\ref{pa}) and (\ref{A}), $\ln(\psi_0)$ is given by
\be
i\left[-\f{(4\Lambda a^2-\lambda^2)^{3/2}}{24\Lambda\kappa}+i\ln\left(\sqrt{\f a{\sqrt\kappa}}\left(\Lambda a^2-\f{\lambda^2}4\right)^{1/4}\right)\right].
\ee
The logarithmic term decays with respect to $S$ as soon as $a\gg\lambda/\sqrt\Lambda$ and in this region $\psi_0\sim e^{iS}$. It is worth noting that to have a self-consistent scheme both relations $l_{in}\gg l_h\sim a$ (which ensures the validity range of the BKL picture) and $l_{in}\gg l_\Lambda$ ($l_\Lambda\equiv1/\sqrt\Lambda$ being the inflation characteristic length) have to hold. (We stress that the inflationary scenario usually takes place if $a\gg l_\Lambda$ \cite{Kolb,KM,iso}). These constraints state the degree of inhomogeneity that is allowed in our model so that the generic Universe can be described, point by point, by a Mixmaster model. At the same time the scale factor $a$ can be considered as a semiclassical variable playing the internal observer-like role for the quantum dynamics of the Universe anisotropies. 

Let us now analyze the hypothesis of a quasi-isotropic potential. A maximum quantum number indicating that the mean value of the quantum Hamiltonian is not compatible with an oscillatory potential has to be found. The expectation value of the Hamiltonian $H_q$ in state $|n\rangle$ is given by
\be\label{eval}
\langle H_q\rangle_n=\f12\left(n+\f12\right)\left(\rho^{-2}+\omega^2\rho^2+\dot\rho^2\right).
\ee   
These values are equally spaced at every instant as for a time-independent harmonic oscillator. The ground state is obtained for $n=0$. The limit of applicability of our scheme relies on an upper bound of the occupation number $n$. Taking into account the expressions for $\omega^2(\tau)$ (\ref{osceq}) and $\rho(\tau)$ (\ref{rho}), the expectation value (\ref{eval}), far from the cosmological singularity, behaves like 
\be
\langle H_q\rangle_n\stackrel{\tau\rightarrow0}\sim\f1{\tau^{5/2}}\left(n+\f12\right).
\ee 
It is then possible to obtain the maximum admissible quantum number $n_{\max}$. The approximation of small anisotropies does not work for a given value of $\beta_\pm$ (for example $\beta_\pm\sim\mathcal O(1)$). Such hypothesis is thus incorrect as soon as $\langle H_q\rangle_n\sim V^\star=V(\beta_\pm\sim\mathcal O(1))$. The value $n_{\max}$ then depends on time $\tau$ and reads $n_{\max}\sim V^\star\tau^{5/2}$. In this way, when a suitablly large configuration of the Universe is taken into account ($\tau\rightarrow0$), not many excited states can be considered. This is not, however, a severe limitation since it is expected that, when the Universe moves away from the classical singularity, the ground state becomes the favored configuration \cite{BM06}.

\section{Concluding remarks} 

In this paper we have shown how a semiclassical isotropization mechanism for a quasi-isotropic inhomogeneous Universe takes place. The wave function of the Universe describes its two physical degrees of freedom (the shape changes $\beta_\pm$), and it is meaningfully interpreted as soon as the isotropic scale factor $a$ plays the observer-like role. This condition is satisfied for large values of the volume, and the dynamics of the anisotropies can be probabilistically interpreted since it describes a small quantum subsystem. The probability density of the possible Universe configurations is stretched over all values of anisotropies near the cosmological singularity. On the other hand, at large scales the probability density is sharply peaked around the closed FRW model. In this way, as the Universe expands from the initial singularity, the FRW state becomes more and more probable.

\section*{Acknowledgments}

M.V.B. thanks ``Fondazione Angelo Della Riccia'' for financial support.

\end{document}